\documentclass[12pt]{article}

\usepackage{amsmath,amssymb, fullpage}
\usepackage{graphicx}
\usepackage[latin1]{inputenc}
\usepackage[T1]{fontenc}
\usepackage{lmodern}
\usepackage[pdfstartview={FitH}]{hyperref}
\usepackage{algorithm}
\usepackage{algorithmic}
\newcommand{\lecture}[4]{
   \pagestyle{myheadings}
   \thispagestyle{plain}
   \newpage
   \setcounter{page}{1}
   \noindent
}
\newtheorem{theorem}{Theorem}
\newtheorem{lemma}{Lemma}

\newtheorem{corollary}[theorem]{Corollary}
\newtheorem{defn}{Definition}

\newtheorem{observation}{Observation}

\def\beq{\begin{eqnarray}}
\def\eeq{\end{eqnarray}}
\def\beqs{\begin{eqnarray*}}
\def\eeqs{\end{eqnarray*}}

\newcommand{\R}{\mathbb{R}}

\newcommand{\A}{\mathbb{A}}
\newcommand{\x}{\mathbf{x}}
\newcommand{\y}{\mathbf{y}}
\newcommand{\z}{\mathbf{z}}
\newcommand{\n}{\mathbf{n}}

\newcommand{\E}{\mathbb{E}}
\newcommand{\p}{\mathbb{P}}
\newcommand{\e}{\mathbf{e}}
\newcommand{\one}{\mathbf{1}}

\newcommand{\la}{\leftarrow}
\def\A{{\mathcal{A}}}
\def\ie{i.\,e.\,}

\title{Geographic Gossip on Geometric Random Graphs via Affine Combinations}
\author{Hariharan Narayanan\\
Department of Computer Science, University of Chicago\\
 {\tt hari@cs.uchicago.edu} }

\begin{document}
\maketitle
\begin{abstract}
In recent times, a considerable amount of work has been devoted to
the development and analysis of gossip algorithms in Geometric
Random Graphs. In a recently introduced model termed ``Geographic
Gossip," each node is aware of its position but possesses no further
information. Traditionally, gossip protocols have always used convex
linear combinations to achieve averaging. We develop a new protocol
for Geographic Gossip, in which counter-intuitively, we use {\it
non-convex affine combinations} as updates in addition to convex
combinations to accelerate the averaging process. The dependence of
the number of transmissions used by our algorithm on the number of
sensors $n$ is $n \exp(O(\log \log n)^2) = n^{1 + o(1)}$. For the
previous algorithm, this dependence was $\tilde{O}(n^{1.5})$.
The exponent 1+ o(1) of our algorithm is asymptotically optimal. Our
algorithm involves a hierarchical structure of $\log \log n$ depth
and is not completely decentralized. However, the extent of control
exercised by a sensor on another is restricted to switching the
other on or off.
\end{abstract}

\section{Introduction}

Geometric Random Graphs have become an accepted model for wireless
ad hoc and sensor networks. Due to applications in distributed
sensing, a significant amount of effort has been directed towards
developing energy efficient algorithms for information exchange on
these graphs. The problem of distributed averaging  has been studied
intensively because it appears in several applications such as
estimation on ad hoc networks, and encapsulates many of the
difficulties faced in asynchronous distributed computation. Let
$v_1, \dots, v_n$ be $n$ points independently chosen uniformly at
random from a unit square in $\R^2$. A Geometric Random Graph $G(n,
r)$ is obtained from these points by connecting any two points
within Euclidean distance $r$. A Gossip Algorithm is an averaging
algorithm that, after a certain number of information exchanges and
updates, leaves each node with a value close to the average of all
the originally held values.

\subsection{Related Work}
There is an extensive body of work surrounding the subject of gossip
algorithms in various contexts. Here, we only survey the results
relevant in a narrow sense to the question under consideration.

Gupta and Kumar \cite{kumar} gave conditions under which $G(n, r)$
is connected with high probability (w.h.p.). It is sufficient that
$r$ scales as $\Omega(\sqrt{\frac{\log n}{n}})$ in order that $G(n,
r)$ be connected with probability greater than $1 - n^{-\Theta(1)}$.

A distributed Gossip Algorithm for arbitrary graphs was presented by
Boyd et al \cite{Boyd}.  In this algorithm, when the clock of a
sensor $s$ ticks, $s$ sends its value $x_s$ to a sensor $v$ chosen
uniformly at random from its neighbors, and receives the value $x_v$
of $v$. Thereafter $s$ and $v$ set their values to $\frac{x_s +
x_v}{2}$. The dependence of the number of transmissions required by
this algorithm on $n$ is $\tilde{O}(n^2)$. The performance was
related to the mixing time of the natural random walk on that graph.
In fact they showed that if the connectivity graph is $G$, the
number of transmissions made in the course of the algorithm is
$\Theta(n T_{mix}(G))$, where $T_{mix}(G)$ is the mixing time of
$G$.

In the standard framework for modeling sensor networks, $n$ sensors
are placed at random on a unit square $\square$ and have a radius of
connectivity $r = \Theta(\sqrt{\frac{\log n}{n}})$. One does not
assume that a sensor possesses any information about its own
location. In this model, the number of transmissions that the best
known algorithm uses is $\tilde{O}(n^2)$ as described
above.\footnote{In using $\tilde{O}$, we ignore polylogarithmic
factors and depending on context, the dependence on parameters other
than $n$.}

A more powerful model was proposed by Dimakis et al
$\cite{wainwright}$, wherein each sensor is aware of its own
location with reference to $\square$ , but possess no further
information. It is mentioned in \cite{wainwright} that this is
reasonable in typical scenarios. With this model, by exploiting
geographic information, they were able to provide an algorithm that
requires $\tilde{O}(n^{1.5})$ transmissions. In their algorithm,
each node exchanges its value with the node nearest to a position
chosen randomly on $\square$, and both nodes replace their values by
the average as in the algorithm of Boyd et al \cite{Boyd}. Rejection
sampling is used to make the distribution roughly uniform on nodes.
The routing takes $\tilde{O}(\sqrt{n})$ hops w.h.p, but since the
mixing time on the complete graph is $O(1)$, one obtains an
algorithm using $\tilde{O}(n^{1.5})$ transmissions, which is an
improvement over \cite{Boyd} by a factor of $\tilde{O}(\sqrt{n})$.

A natural approach to obtaining more efficient algorithms would be
to engage in long-range information exchanges less frequently than
short-range ones. However, it appears that the benefit derived from
an improved mixing time with long-range transmissions more than
compensates for the additional cost in terms of hops for a
long-range routing. Due to this fact, simply altering the
probability distribution with which a node picks targets seems to be
counterproductive.

\subsection{Our Contribution}
An affine combination of two vectors $\mathbf{a}$ and $\mathbf{b}$
has the form $\alpha \mathbf{a} + (1-\alpha) \mathbf{b}$. Unlike the
case of convex combinations, $\alpha$ need not belong to $[0, 1]$.
We introduce counter-intuitive update rules which are {\it affine
combinations} rather than {\it convex combinations} (with
coefficients possibly as large as $\Omega(\sqrt{n})$) to achieve
faster averaging. The total number of transmissions used by the
proposed algorithm in order that the $\ell_2$-distance of the output
from the average diminish by a multiplicative factor of $\epsilon$
w.h.p, is $n\exp(O((\log\log n) \log \log \frac{n}{\epsilon}))$.
When $\epsilon = \exp(n^{\frac{o(1)}{\log \log n}})$ the number of
transmissions is $n^{1+o(1)}$.
 The
exponent $1 + o(1)$ is asymptotically optimal, since every node must
make at least one transmission for an averaging algorithm to work.
Like previous algorithms, ours makes packet exchanges with random
nodes. Due to
 the instability introduced
into the system by the use of non-convex combinations, for the
present analysis to hold, a certain amount of control needs to be
exercised and our algorithm is not truly decentralized. However, the
extent of control exerted by any sensor on another is restricted to
switching the other on or off.

\section{Preliminaries}
The standard model for a sensor network is as follows.
 We assume
that each node or sensor has a clock that is a Poisson process with
rate $1$, and that these processes are independent. This model is
equivalent to having a single clock that is Poisson of rate $n$, and
assigning clock ticks to nodes uniformly at random. We assume that
the time units are adjusted so communication time between any two
adjacent nodes is insignificant in comparison with the length of an
average time slot $n^{-1}$. Our algorithm involves packet forwarding
when two non-adjacent nodes communicate. We shall assume that the
time taken to forward a packet is also insignificant in comparison
with $n^{-1}$, and that a single packet exists in the network in
each time slot w.h.p.. We assume some limited computational power,
which amounts to memory of logarithmic size, and the ability to do
floating point computations.

For our purposes, a Geometric Random Graph is defined in the
following way.
 Let $v_1, \dots, v_n$ be $n$ points independently chosen uniformly at random
from a unit square in $\R^2$. A Geometric Random Graph $G(n, r)$ is
obtained from these points by connecting any two points within
Euclidean distance $r$.

\subsection{Problem Statement}
Let node $v_i$ for  $i = 1, \dots, n$ hold a value $x_i(t)$ at the
$t^{th}$ global clock tick, the initial values being $x_i(0)$.
Without loss of generality, we assume $\overline{\x(0)} = 0$. Given
$\epsilon, \delta > 0$,  the task is to design an algorithm
such that $\|\x(t)\| < \epsilon \|\x(0)\|$  for all possible choices
of $\x(0)$  with probability $> 1 - \delta$. The cost of the
algorithm is the expected number of transmissions made until $t$.

In the rest of the paper, we shall make the standard assumption that
the radius of connectivity $r(n) = \Theta(\sqrt{\frac{\log n}{n}})$
(eg \cite{wainwright}.) Under this assumption, the probability of
the graph $G(n, r)$ being disconnected is $\Omega(n^{-O(1)})$, for
an appropriate constant $a$.  As a consequence, it is not possible
to drive $\delta$ below $n^{-O(1)}$. For this reason, in the
analysis, we shall assume that $\delta = n^{-O(1)}$. On the other
hand $\epsilon$ can be made arbitrarily small by running the
averaging algorithm for a sufficiently long interval of time. In
this paper, we shall assume that $\log \frac{1}{\epsilon} =
n^{\frac{o(1)}{\log \log n}}$. This does not allow $\epsilon$ to be
exponentially small but permits it to be the reciprocal of a
quasipolynomial. A sufficiently large constant $a$ will appear in
the parameters of our algorithm described later.When we use the term
{\it high probability}, we shall mean with probability $1 -
n^{-\Theta(1)}$.

\section{Overview of Algorithm}

Let $\square$ be the unit square in which the $n$ sensors are
randomly placed. Let the initial values carried by sensors be
$x_i(0)$, for $i = 1$ to $n$.  We consider a partition of $\square$
into $\sim n^{1/2}$ smaller squares $\square_i$. Let $\square_i$
contain $\#(\square_i)$ sensors. Let $time(n)$ represent the
expected number of transmissions until $\|\x(t)\| \leq \epsilon
\|\x(0)\|$  w.h.p., where $\epsilon$ is some function of $n$ that we
shall not investigate at the moment. Suppose that we had a ``nearly
perfect" averaging protocol $\A$ on the smaller squares $\square_i$,
\ie when $\A$ is run on each square, after $t = time_{\A}(\sqrt n)$
transmissions, within $\square_i$ the values are for practical
purposes equal to the the average of the original values. That is,
$$(\forall i) (\forall s \in \square_i) x_s(t) \backsimeq \frac{\sum\limits_{s \in
\square_i} x_s(0)}{\#(\square_i)}.$$
\begin{defn}
For each square $\square_i$, let $s(\square_i)$ be the sensor
closest to the center of $\square_i$.
\end{defn}
This can be determined by each square, using a constant number of
transmissions w.h.p.

The $s(\square_i)$ exchange values among themselves by Greedy
Geographic Routing (see \cite{wainwright}).

Consider the following protocol. Suppose that $\A$ has been run on
each subsquare of the form $\square_i$ independently, and the values
carried by the nodes within $\square_i$ are all equal. When
$s(\square_i)$ becomes active,
 the following round takes place.
\begin{enumerate}
\item $s_i :=s(\square_i)$ picks a square $\square_j$ uniformly at
random. $s_i$ geographically routes a packet with its value to $s_j
:= s(\square_j)$.

\item $s_j$ routes its own value to $s_i$ by greedy
geographic routing.
\item $x_{s_i} \la x_{s_i} + \frac{2\sqrt{n}}{5}(x_{s_j} - x_{s_i})$.
\item  $x_{s_j} \la x_{s_j} + \frac{2\sqrt{n}}{5}(x_{s_i} - x_{s_j})$.
\item $\A$ is independently run on $\square_i$ (the process being activated by $s_i$ by switching certain nodes on)
and on $\square_j$ (initiated by $s_j$ similarly).
\item $\A$ is ended on square $\square_i$ by $s_i$ (by turning certain nodes off), and
$\A$ is ended on $\square_j$ by $s_j$ (by switching certain nodes
off.)
\end{enumerate}

Now, let $z_i(t) := \sum\limits_{s \in \square_i} x_s(t)$. Without
loss of generality, we assume that $\sum_i{x_i} = 0$, since this
only adds a constant offset and does not affect the rate of
convergence. An application of the Chernoff Bound tells us that
$(\forall i)\left| \frac{\#(\square_i)}{\sqrt n}-1\right| <
\frac{1}{10}$ w.h.p . If we examine the evolution of $\z$, we see
that after a round of the kind described above

\begin{itemize}
\item $z_i(t) = (1-\alpha_i)z_i(t-1)
+ \alpha_jz_j(t-1)$
\item $z_j(t) = (1-\alpha_j) z_j(t-1) + \alpha_i z_i(t-1)$
\end{itemize}
where $\forall i,  \alpha_i \in (\frac{1}{2}, \frac{1}{3})$. From
Lemma~\ref{l:1}, it follows that

$\E[\|\z(t)\|^2] < (1-\frac{1}{2\sqrt{n}})^t \|\z(0)\|^2$. Roughly
speaking after $O(\sqrt{n} \log(\frac{n}{\epsilon}))$ of these
steps, we have a distribution $\x(t')$ such that $\|\x(t')\| <
\epsilon \|\x(0)\|$.

Each geographical routing mentioned above takes $O(\sqrt n)$
transmissions w.h.p (see \cite{wainwright}). Also, each process of
initiating or ending $\A$ on a square $\square_i$ takes
$O(\sqrt{n})$ transmissions.

So, the total number of transmissions with $n$ nodes $time(n)$
satisfies a recurrence of the form: $$time(n) \backsimeq
O\left(\sqrt n \log(\frac{n}{\epsilon}) ( time_\A(\sqrt n) + O(\sqrt
n))\right).$$ Ignoring the dependence on $\epsilon$, it would allows
us to recursively define the algorithm $\A$ on $\square$, for which
$time_\A(n) = n \exp(O(\log\log n)^2).$

\section{Description of the Algorithm}
\subsection{Notation}\label{s:1}
The square $\square$ is partitioned into $n_1$  subsquares
$\square_i$, where $n_1$ is the nearest integer to $\sqrt{n}$ that
is the square of an even number. For a square $\square_{i_1\dots
i_r}$, let $\E_\#\square_{i_1\dots i_r}$ denote the expected number
of sensors within $\square_{i_1\dots i_r}$. Then, while
$\E_\#\square_{i_1\dots i_r} > (\log n)^8$,

the square $\square_{i_1\dots i_r}$ is partitioned into $n_{r+1}$
subsquares $\square_{i_1\dots i_{r+1}}$, where $n_{r+1}$ is the
nearest integer to $\sqrt{\E_\#\square_{i_1\dots i_r}}$ that is the
square of an even number. Let $$\ell := 1 +
\sup\limits_{\square_{i_1\dots i_r}} r,$$ \ie the number of levels
in this recursion. Given a square $\square_{<i>}$, let
$s(\square_{<i>})$ denote the sensor nearest to its center. By our
construction, these centers are well separated, and any sensor has
this property with respect to at most one square w.h.p.. We shall
denote
this by $\square(s)$. 
We assign a Level to each node by the
following rule: If $s = s(\square_{i_1\dots i_r})$, $s$ has level
$\ell - r$. These nodes are have Levels $1, \dots, \ell$. There is a
single root node at Level $\ell$, namely $s(\square)$. The nodes at
Level $0$ are the nodes not of the form $s(\square_{i_1\dots i_r})$.
In the informal discussion earlier, we did not concern ourselves
with the error in the averaging carried out on subsquares
$\square_i$. However, these errors propagate up the hierarchy
rapidly, and hence it is necessary to obtain results with greater
accuracy in smaller squares. Thus we define the desired accuracy
recursively. Let $\epsilon_r$ be the accuracy for the averaging
process in a square $\square_{i_1 \dots i_{r-1}}.$ Lemma~\ref{l:2}
tells us that it is sufficient to take $\epsilon_r$, to be
$\frac{\epsilon_{r-1}}{\text{poly}(n)}$ for a polynomial of
sufficiently large degree.

 Let $\epsilon_0 =
\epsilon$, $\delta_0 = \delta$. We recursively define
$\epsilon_{r+1} := \frac{\epsilon_r}{25 n^{\frac{7}{2} + a}}$ and
$\delta_{r+1} = \frac{\delta_r}{n_r^{2a}}$.

We define $time(n, \ell-1, \epsilon_r, \delta_r)$ to be $\left((\log
\frac{n}{\epsilon_{\ell-1}})
\log(\delta_{\ell-1}^{-1})\right)^{16}$. Thereafter, we define
$time(n, r-1, \epsilon_{r-1}, \delta_{r-1}) := time(n, r,
\epsilon_r, \delta_r) n^a \left(\log (\frac{n_r}{\epsilon_r})\log
(\delta_r^{-1})\right)^{16}.$

Let  $s \in \square_{i_1\dots i_{\ell-1}}$.

\subsection{The Protocol}
Every node $s$ has two states, a $local.state$ and a $global.state$,
both of which are initially $= off$, but can also take the value
$on$. Each node $s$ possesses a private counter $counter(s)$. During
initialization, the $global.state$ of $s(\square)$ is set to $on$
but every other $global.state$ is $0$. The $local.state$ of {\it
all} nodes is set to $off$ at this juncture.

Let us suppose that the clock of $s$ ticks. We describe the protocol
followed by it below. We consider two cases. If $s$ is at Level $0$,
it obeys the following protocol: \{
\begin{enumerate}
\item If $local.state(s) = on$\\ $Near(s)$;
\end{enumerate}
\}

$Near(s)$\{
\begin{enumerate}
\item $s$ picks an adjacent node $v$ contained in $\square_{i_1\dots i_{\ell-1}}$
uniformly at random.
\item $s$ sets $x_s(t+1) = \frac{x_s(t) + x_v(t)}{2};$\\
      $v$ sets $x_v(t+1) = \frac{x_s(t) + x_v(t)}{2};$
\end{enumerate}
\}

We next describe the protocol if $s$ is at a Level greater than $0$.
The subroutine $Near$ is the same as above.
Let $\square(s)=: \square_{i_1 \dots i_r}.$\\
\{
\begin{enumerate}
 \item If $global.state(s) = on$ 

 \begin{enumerate}
 \item  If $counter(s) = 0$ $Activate.square(s);$
 \item With probability $n^{-a} time(n, r, \epsilon_r,
 \delta_r)^{-1}$
 \begin{itemize}
 \item $Far(s)$;
 \item $counter(s) \la 0$;
 \end{itemize}

\end{enumerate}

\item If  $local.state(s) = on$ \\$Near(s);$

\item If $counter(s) \geq time(r, n, \epsilon_r, \delta_r)$
$Deactivate.square(s);$\\
Else $counter(s) \la counter(s) + 1;$

\end{enumerate}
\}

${Far(s)}$\{
\begin{enumerate}
\item $s$ picks a square $\square_{i_1'\dots i_r'} \not\ni s$ uniformly at
random. Let $s' := s(\square_{i_1'\dots i_r'})$ . Node $s$ routes
its value to $s'$ geographically.


\item $x_s(t+1) = x_s(t) + \frac{2}{5}(\E_\#\square_{i_1\dots i_r} x_{s'}(t) -
\E_\#\square_{i_1 \dots i_r}x_s(t))$.
\item $s'$ sends  back to a packet with its value $x_{s'}(t)$ to  $s$ by greedy
geographic routing.
\item Node $s$ computes $x_s(t+1) = x_s(t) + \frac{2}{5}(\E_\#\square_{i_1\dots
i_r}x_{s'}(t) - \E_\#\square_{i_1\dots i_r} x_s(t))$.
\item $counter(v) \la 0$.
\end{enumerate}\}

$Activate.square(s)$\{
\begin{enumerate}
\item If $s \in $ Level $1$, send packets to each node $s'$ in
$\square(s)$ setting $local.state(s') \la on$ by flooding.
\item If $s \in $ Level $i>1$, send packets to each Level $i-1$ node $s'$ in
$\square(s)$ by greedy geographic routing, setting $global.state(s')
\la on$.
\end{enumerate}
\}

$Deactivate.square(s)$\{
\begin{enumerate}
\item If $s \in $ Level $1$, send packets to each node $s'$ in
square($s$) setting $local.state(s') \la off$ by flooding.
\item If $s \in $ Level $i>1$, send packets to each Level $i-1$ node $s'$ in
$\square(s)$ by greedy geographic routing, setting $global.state(s')
\la off$.
\end{enumerate}
\}
\section{Analyzing the number of Transmissions}
Let $H(n, r, \epsilon_r, \delta_r)$ denote the number of
transmissions used in our protocol in one round of
$\square_{i_1\dots i_r}$, in order to diminish the variance (of the
values carried by sensors in $\square_{i_1\dots i_r}$) by a factor
$\epsilon_r$, with probability $1-\delta_r$.



\begin{observation}\label{l:red}
 In one round, \ie the duration
between $s$ activating $\square(s) := \square_{i_1\dots i_r}$ and
deactivating $\square(s)$, the number of long-range packet exchanges
between sensors of the kind $s(\square_{i_1\dots i_r i_{r+1}})$ is
$\Theta\left(\tilde{n} \log(\frac{\tilde{n}}{\epsilon_r})\right)$
w.h.p, where $$\tilde{n} = \frac{\E_{\#}[\square_{i_1\dots
i_r}]}{\E_{\#}[\square_{i_1 \dots i_r i_{r+1}}]}.$$
\end{observation}
  Each of these
involves $O(\sqrt{\E_{\#}[\square(s)]}) \tilde{n}$ hops w.h.p (see
\cite{wainwright}). Therefore the total number of transmissions here
is $O\left(\tilde{n}^2 \log(\frac{\tilde{n}}{\epsilon_r})\right)$
w.h.p.

Each of these long-range packet exchanges is followed by a period of
averaging within the involved subsquares, and this takes $H(n, r+1,
\epsilon_{r+1}, \delta_r) = \Omega(\tilde{n})$ transmissions. Thus
we have the recurrence \beqs \label{e:1} H(n, r, \epsilon_r,
\delta_r) & = & O\left((H(n, r+1, \epsilon_{r+1}, \delta_{r+1}) +
\tilde{n})\tilde{n}\log(\frac{\tilde{n}}{\epsilon_r})\right)\\
 & = &  O\left(H(n, r+1, \epsilon_{r+1}, \delta_{r+1})\tilde{n}\log(\frac{\tilde{n}}{\epsilon_r})\right).
 \eeqs

As mentioned in subsection~\ref{s:1}, we let
 $\epsilon_0 =
\epsilon$, $\delta_0 = \delta$ and  recursively define
$\epsilon_{r+1} := \frac{\epsilon_r}{25 n^{7/2}}$ and $\delta_{r+1}
= \frac{\delta_r}{n_r^2}$.
 For these parameters,
$\delta_r = \Omega(\frac{1}{\text{poly}(n)})$, since $\delta_0 =
\Omega(\frac{1}{\text{poly}(n)})$ and the $\tilde{n}$ telescope.
$\epsilon_r = \epsilon_0\Omega{n^{-O(\log\log n)}}$ since $\ell \sim
\log \log n$.
 Now, the smallest squares that we create have $O(\text{polylog}n)$
 sensors each w.h.p. Since the ordinary averaging that we do there
 (described by the procedure "Near(s)") has an averaging time that is
 quadratic \cite{Boyd, Boyd2},
$H(n, \ell, \epsilon_{\ell}, \delta_\ell) =
\Omega(\text{polylog}(\frac{n}{\epsilon_\ell}))$. And so using the
recurrence for $H$ and telescoping, we see that the total number of
transmissions is
 \beqs H(n, 0, \epsilon_0, \delta_0) &=& \left(H(n,
\ell, \epsilon_{r+1}, \delta_{r+1})\right)\prod_r
\left\{\frac{\E_{\#}[\square_{i_1\dots i_r}]}{\E_{\#}[\square_{i_1
\dots i_r i_{r+1}}]} \log \frac{n}{\epsilon_r}\right\}\\
& = & n (\log \frac{n}{\epsilon})^{O(\log \log n)}. \eeqs This is
$n^{1+o(1)}$ if $\epsilon = \exp(-n^{\frac{o(1)}{\log \log n}})$,
and $\delta = n^{-O(1)}$.

\section{Notes on Correctness}
In the algorithm proposed in this paper, each square $\square(s)$
has a certain latency, which is the averaging time restricted to
that square. In order for our algorithm to be correct, we require
that $\square(s)$ be undisturbed by the long-range exchanges that
$s$ is involved in, during this period. This is not a condition that
can be imposed without the long-range exchanges of $s$ losing their
i.i.d property, which is crucial in our analysis of convergence. In
order to retain this, and have an algorithm that is successful w.h.p
we have set the rates at which long-range exchanges of $s$ occur to
be lower than the inverse of the latency by a factor $n^a$. As a
consequence, w.h.p, in the course of the entire algorithm, there are
no long-range transmissions made by any node $s$ while $\square(s)$
is active. The only issue that we have not dealt with in detail is
of showing that our choice of errors $\epsilon_r$ achieves the
desired end. This follows from Lemma~\ref{l:2} interpreted as
follows: The nodes $i$  represent {\it subsquares} $\square_{i_1
\dots i_r i_{r+1}}$ of $\square_{i_1 \dots i_r}$ and the $y_j(t)$
for different $j$ represent  the {\it sum} of the values held by the
nodes in a subsquare $\square_{i_1 \dots i_r j}$ after $t$ long
distance transmissions between subsquares since the activation of
$\square_{i_1 \dots i_r}$. We set $\epsilon := {\epsilon_{r+1}}
\|\x(0)\|$. The perturbations $n(t)$ represent the errors generated
from imperfect averaging within these subsquares.

\section{Concluding Remarks}
We introduced {\it non-convex affine combinations}, in our averaging
protocol in order to accelerate Geographic Gossip in Geometric
random graphs. The number of transmissions used in the course of our
protocol is $n^{1+o(1)}$. This exponent is asymptotically optimal.
Our algorithm, unlike the previous one in \cite{wainwright} is not
completely decentralized. However as far as we can see, this is not
a necessary feature associated with the use of affine combinations.



\section{Future Directions}
It would be interesting to study whether affine combinations can be
used to develop a completely decentralized algorithm for Geographic
Gossip that is also energy efficient.

\appendix
\section{Appendix}
 Let $K_n$ be the complete graph on $n$ vertices $\{1, \dots,
n\}.$ $\forall i,$ let $\alpha_i \in (\frac{1}{3}, \frac{1}{2}).$ At
time $t \geq 0$, for $i = 1, \dots, n$, let node $i$ hold the value
$x_i(t)$. Consider the following update rule. If the $t^{th}$ clock
tick belongs to node $i$, then, $i$ chooses a node $j$ uniformly at
random, and the following update occurs:

\begin{itemize}\label{update}
\item $x_i(t) = (1-\alpha_i) x_i(t-1) + \alpha_j x_j(t-1) .$
\item $x_j(t) = (1-\alpha_j) x_j(t-1) + \alpha_i x_i(t-1).$
\end{itemize}

\begin{lemma}\label{l:1}
$\E[\x(t)^T \x(t)] < (1-\frac{1}{2n})^t \x(0)^T \x(0)$.
\end{lemma}
{\bf Proof:}
Let the update rule for $\x(t)$ be given by $A(t-1)$, \ie \, $\x(t)
= A(t-1)\x(t-1)$. Note that $A(t-1) = I - (\alpha_i \e_i - \alpha_j
\e_j)(\e_i^T - \e_j^T)$, if the $i^{th}$ vector of the standard
basis is denoted by $\e_i$.

\beqs\label{eq1}
\E[\x(t)^T \x(t)|\x(t-1)] & = & \E[\x(t-1)^T A(t-1)^T A(t-1) \x(t-1)|\x(t-1)]\\
               & = & \x(t-1)^T \E[A(t-1)^T A(t-1)] \x(t-1).
               \eeqs
Let $\alpha_i \e_i - \alpha_j \e_j = \mathbf{\alpha}_{ij}$ and $\e_i
- \e_j = \e_{ij}$. Then, $\E[A(t-1)^T A(t-1)] = \E[(I -
\e_{ij}\alpha_{ij}^T)^T(I - \e_{ij}\alpha_{ij}^T)]$.

Let $E_{ij}$ denote the $n \times n$ matrix whose $ij^{th}$ entry is
$1$ and every other entry is $0$.

Then, by expanding, one finds that \beqs \E[A(t-1)^T A(t-1)] & = & I
+ \sum_i \frac{(1-2\alpha_i)^2 -1}{n} E_{ii} + \sum_{i \neq j}
\frac{(1 - (1-2\alpha_i)(1-2\alpha_j)) E_{ij}}{n(n-1)}\\
& = & I (1 - \frac{1}{n-1}) + \frac{\mathbf{1}\mathbf{1}^T}{n(n-1)}
- \frac{(\mathbf{1}-2\mathbb{\alpha})(\mathbf{1}-2\alpha)^T}{n(n-1)}
+ \sum_i\frac{(1-2\alpha_i)^2E_{ii}}{n-1}. \eeqs An application of
the formula for $\E[\x(t)^T \x(t)|\x(t-1)]$, now gives us the
following:

\beq \label{expr} \E[x(t)^T x(t) | x(t-1)] & = & \E[x(t-1)^TA(t-1)^TA(t-1)x(t-1)|x(t-1)]\\
                              & = & x(t-1)^T\E[A(t-1)^TA(t-1)]x(t-1)
                            \eeq
We know that $\forall i,  1-2\alpha_i \in (0, \frac{1}{3})$.

Let us upper bound $x(t-1)^T\E[A(t-1)^TA(t-1)]x(t-1)$ using the the
expression for $\E[A(t-1)^T A(t-1)]$ derived earlier.
$$x(t-1)^T I (1 - \frac{1}{n-1}) x(t-1) = (1 -
\frac{1}{n-1})\|x(t-1)\|^2,$$
$$\frac{x(t-1)^T \mathbf{1} \one^T x(t-1)}{n-1} = 0,$$
$$- \frac{x(t-1)^T (\mathbf{1} - 2\alpha)(\one^T - 2\alpha^T)
x(t-1)}{n(n-1)} \leq 0 $$ and,
$$x(t-1)^T \left(\sum_i \frac{(1-2\alpha_i)^2
E_{ii}}{n-1}\right)x(t-1)  \leq \frac{\|x(t-1)\|^2}{9(n-1)}.$$

Adding up the above inequalities, $$\E[x(t)^Tx(t)|x(t-1)] \leq
\left(1 - \frac{8}{9(n-1)}\right) x(t-1)^T x(t-1).$$ As a
consequence,
$$\E[\|x(t)\|^2 \, | \, x(t-1)] < \left(1 - \frac{1}{2n}\right) \|x(t-1)\|^2.$$
Successively conditioning on $x(t-2), \dots, x(0)$, we see that
$$\E[\|x(t)\|^2] < \left(1 - \frac{1}{2n}\right)^t \|x(0)\|^2.$$
This proves the lemma.{\hfill $\Box$}

An application of Markov's inequality gives us the following
corollary.
\begin{corollary}\label{c:1}
$$\p\left(\|x(t)\| > \epsilon \|x(0)\|\right) \leq \epsilon^{-2}\left(1 -
\frac{1}{2n}\right)^t.$$
\end{corollary}
{\bf Proof:}
\beqs \p\left(\|x(t)\| > \epsilon \|x(0)\|\right) &=& \p\left(\frac{\|x(t)\|^2}{\|x(0)\|^2} > \epsilon^2 \right)\\
                                                &\leq& \epsilon^{-2}\E\left(\frac{\|x(t)\|^2}{\|x(0)\|^2}\right) {\hfill (\text{Markov's inequality})} \\
                                                &\leq& \epsilon^{-2}\left(1 - \frac{1}{2n}\right)^t
\eeqs    {\hfill $\Box$}

An application of Markov's inequality gives us the following
corollary.
\begin{corollary}\label{c:1}
$$\p\left(\|x(t)\| > \epsilon \|x(0)\|\right) \leq \epsilon^{-2}\left(1 -
\frac{1}{2n}\right)^t.$$
\end{corollary}

We now consider a modified update rule, and prove a lemma similar to
Lemma~\ref{l:1}.

Let $K_n$ be the complete graph on $n$ vertices $\{1, \dots, n\}.$
$\forall i,$ let $\alpha_i \in (\frac{1}{3}, \frac{1}{2}).$ At time
$t \geq 0$, for $i = 1, \dots, n$, let node $i$ hold the value
$x_i(t)$. Let $n(0), n(1), \dots$ be a sequence of real numbers.
Consider the following update rule. If the $t^{th}$ clock tick
belongs to node $i$, then, $i$ chooses a node $j$ uniformly at
random, and the following update occurs:

\begin{itemize}\label{update}
\item $y_i(t) = (1-\alpha_i) y_i(t-1) + \alpha_j y_j(t-1) + n(t-1).$
\item $y_j(t) = (1-\alpha_j) y_j(t-1) + \alpha_i y_i(t-1) - n(t-1).$
\end{itemize}

\begin{lemma}\label{l:2}
Suppose that for each $t$, $|n(t)| < \epsilon$, and that $a
> 0$. Then,
$$\p\left[\|\y(t)\| > n^{\frac{a}{2}}\left((1-\frac{1}{2n})^{t/2}\|\y(0)\| + 8\sqrt{2} n^{3/2}
\epsilon \right)\right] \leq \frac{5}{n^a}.$$
\end{lemma}
{\bf Proof:}
 $\y(t) = A(t-1)\y(t-1) + \n(t-1)$,
where $A(t) = I - (\alpha_i \e_i - \alpha_j \e_j)(\e_i^T - \e_j^T)$,
and $\n(t-1) = n(t-1)(\e_i - \e_j).$ Let $\x(0) = \y(0)$, and let
the $\x(t)$ satisfy $\x(t+1) = A(t)\x(t)$ as in Lemma~\ref{l:1}. We
observe that
$$\y(1) = \x(1) + \n(0)$$ and more generally,
$$ \y(t+1) =  \x(t+1) + \n(t) + \sum_{i=0}^{t-1} A(t)A(t-1)
\dots A(i+1) \n(i) .$$ An application of the triangle inequality now
gives us
$$ \|\y(t+1)\| \leq  \|\x(t+1)\| + \|\n(t)\| + \sum_{i=0}^{t-1} \|A(t)A(t-1)
\dots A(i+1) \n(i)\| .$$ Our approach to proving this Lemma is to
upper bound each term in the right hand side.
\begin{observation}\label{o:1}
\beqs \p\left[\|\x(t)\|
 >  (1-\frac{1}{2n})^{t/2} n^{a/2} \|x(0)\|\right] & \leq &
\left((1-\frac{1}{2n})^{t/2}
n^{a/2}\right)^{-2}\E\left(\frac{\|x(t)\|^2}{\|x(0)\|^2}\right)\\
& \leq & \left((1-\frac{1}{2n})^{t/2} n^{a/2}\right)^{-2} (1 -
\frac{1}{2n})^t \\
& = & \frac{1}{n^a}.\eeqs
\end{observation}
The above inequalities follow from Lemma~\ref{l:1} and
Corollary~\ref{c:1}. We shall now upper bound the other terms as
well with high probability. Using Corollary~\ref{c:1} \beqs
\p\left[\frac{\|A(t-1) \dots A(i) \n(i-1)\|}{\|\n(i-1)\|} >
(1-\frac{1}{2n})^{\frac{t-i}{4}} n^{\frac{a+1}{2}}\right] & \leq &
((1-\frac{1}{2n})^{\frac{t-i}{4 }} n^{\frac{a+1}{2}})^{-2}\left(1 -
\frac{1}{2n}\right)^{t-i}\\
& = & n^{-(a+1)}(1-\frac{1}{2n})^{\frac{t-i}{2}}. \eeqs

However,
$$\sum_{i=1}^{t-1} (1-\frac{1}{2n})^{\frac{t-i}{2}} n^{-(a+1)} <
\frac{4}{n^a}$$ and so,
$$ \p\left[\exists_i \left\{\frac{\|A(t-1) \dots A(i)
\n(i-1)\|}{\|\n(i-1)\|}
> (1-\frac{1}{2n})^{\frac{t-i}{4}} n^{\frac{a+1}{2}}\right\}\right] \leq
\frac{4}{n^a}.$$

We next observe that $$\sum_{i \leq t}
(1-\frac{1}{2n})^{\frac{t-i}{4}} n^{\frac{a+1}{2}} < 8
n^\frac{a+3}{2}.$$ As a consequence we have

\begin{observation}
$$ \p\left[\sum_i \frac{\|A(t-1) \dots A(i)
\n(i-1)\|}{\|\n(i-1)\|}
> 8 n^\frac{a+3}{2}\right] \leq \frac{4}{n^a}.$$
\end{observation}


Once we put the above two observations together and note that
$(\forall i) \sqrt{2} \epsilon \geq \|\n(i)\|$, an application of
the union bound gives
$$ \p\left[\|\y(t)\| > n^{\frac{a}{2}}\left((1-\frac{1}{2n})^{t/2}\|\y(0)\|+
8 \sqrt{2} n^{3/2} \epsilon \right)\right]  \leq \frac{5}{n^a}.$$
{\hfill $\Box$}

\begin{thebibliography}{50}

\bibitem{Boyd}
S.~Boyd, A.~Ghosh, B.~Prabhakar, and D.~Shah.
\newblock Gossip algorithms : Design, analysis and applications.
\newblock In {\em Proceedings of the 24th Conference of the IEEE Communications
  Society (INFOCOM 2005)}, 2005.

\bibitem{Boyd2}
S.~Boyd, A.~Ghosh, B.~Prabhakar, and D.~Shah.
\newblock Mixing Times for Random Walks on Geometric Random Graphs.
\newblock SIAM ANALCO 2005.

\bibitem{car}
S. ~Carruthers, V. ~King.
\newblock Connectivity of Wireless Sensor Networks with Constant
Density.{\em ADHOC-NOW, 2004}, 149-157
\newblock

\bibitem{kumar}
P.~Gupta and P.~Kumar.
\newblock The capacity of wireless networks.
\newblock {\em IEEE Transactions on Information Theory}, 46(2):388--404, March
  2000.

\bibitem{wainwright}
 A. ~Dimakis, A. ~Sarwate, M. ~Wainwright.
 \newblock Geographic gossip: efficient aggregation for sensor
 networks.
 \newblock In {\em Proceedings of the fifth international conference on information processing in sensor networks (IPSN)}, 2006.

\bibitem{Karp}
R.~Karp, C.~Schindelhauer, S.~Shenker, and B.~V\"{o}cking.
\newblock Randomized rumor spreading.
\newblock In {\em Proc. IEEE Conference of Foundations of Computer Science,
  (FOCS)}, 2000.

\bibitem{k1}
D.~Kempe, J.~Kleinberg, A.~Demers.
 \newblock Spatial gossip and
resource location protocols.
 \newblock in {\em Proc. 33rd ACM
Symposium on Theory of Computing,} 2001.

\bibitem{k2}
D. ~Kempe, J. ~Kleinberg.
\newblock Protocols and Impossibility
Results for Gossip-Based Communication Mechanisms.
 \newblock In
{\em Proc. 43rd IEEE Symposium on Foundations of Computer Science,}
2002.

\bibitem{MoskAoyama}
D.~Mosk-Aoyama and D.~Shah.
\newblock Information dissemination via gossip: Applications to averaging and
  coding.
\newblock http://arxiv.org/cs.NI/0504029, April 2005.

\bibitem{MR95}
R.~Motwani and P.~Raghavan.
\newblock {\em Randomized Algorithms}.
\newblock Cambridge University Press, Cambridge, 1995.

\bibitem{Penrose}
M.~Penrose.
\newblock {\em Random Geometric Graphs}.
\newblock Oxford studies in probability. Oxford University Press, Oxford,
2003.

\bibitem{Xiao}
L.~Xiao, S.~Boyd, and S.~Lall.
\newblock A scheme for asynchronous distributed sensor fusion based on average
  consensus.
\newblock In {\em 2005 Fourth International Symposium on Information Processing
  in Sensor Networks (IPSN)}, 2005.
\end{thebibliography}
\end{document}